\documentstyle[11pt,epsf]{article}
\input{psfig}

\newtheorem{definition}{\bf Definition}

\newtheorem{example}{\bf Example}
\newtheorem{remark}{\bf Remark}

\newcommand{\lea}{\stackrel{+}{<}}
\newcommand{\gea}{\stackrel{+}{>}}
\newcommand{\eqa}{\stackrel{+}{=}}

\date{}

\title{Applying MDL to Learning Best Model Granularity\thanks{Parts of this work were
presented at IJCAI, 1989, and at ESANN, 1994.}}
\author{
Qiong Gao\thanks{Chinese Academy of Sciences,  Institute of Botany, Beijing, China.}\\
Chinese Academy of Sciences\\
\and
Ming Li\thanks{Supported
in part by
ONR Grant N00014-85-K-0445
and ARO Grant DAAL03-86-K-0171
at Harvard University, by NSERC operating grant OGP-0036747 
at York University, and by
NSERC operating grant OGP-046506 at the University of Waterloo.
Address: Computer Science Department,
University of California,
Santa Barbara, CA 93106, USA---On leave from the University of
Waterloo, Waterloo, Canada. Email: mli@cs.ucsb.edu}\\
University of California, Santa Barbara\\
\and
Paul M.B. Vit\'{a}nyi\thanks{Partially supported by the European Union 
through 4th Framework NeuroCOLT II Working Group
EP 27150, the 5th Framework QAIP Project IST-1999-11234, 
the 5th Framework Network of Excellence
QUIPROCONE IST-1999-29064, the COLORET Working Group,
the European Science Foundation,
and by an NSERC
International Scientific Exchange Award ISE0125663.
Address: Centrum voor Wiskunde en Informatica,
Kruislaan 413, 1098 SJ Amsterdam, The Netherlands. Email: paulv@cwi.nl}\\
CWI and University of Amsterdam\\}

\begin{document}
\maketitle
\begin{abstract}
The Minimum Description Length (MDL)
principle is 
solidly based on a provably ideal method of
inference using Kolmogorov complexity. We test how the theory
behaves in practice on a general problem in model selection: 
that of learning the best
model granularity.
The performance of a model depends critically
on the granularity, for example
the choice of precision of the parameters. Too high precision
generally involves modeling of accidental noise and too low precision
may lead to confusion of models that should be distinguished.
This precision is often determined ad hoc. 
In MDL the best model is the one that most compresses a two-part code
of the data set: this embodies ``Occam's Razor.''
In two quite different experimental settings the theoretical
value determined using MDL coincides with the best value found experimentally.
In the first experiment the task
is to recognize isolated
handwritten characters in one subject's handwriting,
irrespective of size and orientation.
Based on a new modification of elastic matching, using multiple prototypes
per character, 
 the optimal prediction rate is predicted
for the learned parameter (length of sampling interval) considered
most likely by MDL, 
which is shown to coincide
with the best value found experimentally.
In the second experiment the task is to
model a robot arm with two degrees
of freedom using a three layer feed-forward neural network where 
we need to determine the number of nodes in the hidden layer giving
best modeling performance. The optimal model (the one that extrapolizes best
on unseen examples) is
predicted for the number of nodes in the hidden layer considered
most likely by MDL, which again is found to coincide with the
best value found experimentally.
\end{abstract}

\section{Introduction}
It is commonly accepted that all learning involves compression
of experimental data in a compact `theory' , `hypothesis', 
or `model' of the phenomenon under investigation. In \cite{ViLi99,LiVi89a}
the last two authors analysed the theory of such approaches 
related to shortest effective description
length (Kolmogorov complexity). The question arises
whether these theoretical insights can
be directly applied 
to real world problems. Selecting models on the basis
of compression properties ignores the `meaning' of
the models. Therefore we should aim at optimizing a model
parameter that has no direct semantics, such as the 
precision at which we represent the other parameters: too high
precision causes accidental noise to be modeled as well, too low
precision may cause models that should be distinct to be confused.
In two quite different experimental settings the theoretically predicted
values are shown to coincide with the best values found experimentally.

In general, the performance of a model for a 
given data sample depends critically
on what we may call the ``degree of discretization'' or the 
``granularity'' of the model:
the choice of precision of the parameters, the number of
nodes in the hidden layer of a neural network, and so on. 
The granularity is often determined ad hoc. Here we give
a theoretical method to determine
optimal granularity based on an application
of the Minimum Description Length (MDL) principle
in supervised learning. The cost of describing a set of data
with respect to a particular model is 
the sum of the lengths of the model description
and the description of the data-to-model error.
According to MDL the best model is one with minimum cost,
that is, the model that explains the data in the most concise way.

The first two authors, in \cite{GL89}, 
carried out an experiment in on-line learning to recognize 
isolated alphanumerical characters
written in 
one subject's handwriting,
irrespective of size and orientation.
Some novel features here are the use of multiple prototypes
per character, and the use of the MDL principle to choose
the optimal feature extraction interval.
It is satisfactory that in this case
the general learning theory can
without much ado be applied to obtain the best sampling rate.
We believe that the same method is applicable
in a wide range of problems. To obtain evidence for this assertion, in
\cite{BKV94} the third author with G. te Brake and J. Kok applied
the same method to modeling a robot-arm.

The genesis of this work is not rooted in traditional approaches
to artificial intelligence (AI), but rather on
new exciting general learning theories which have developed out 
from the computational complexity theory \cite{Va84,Va84a},
statistics and descriptional (Kolmogorov)
complexity \cite{Ri78}. These new theories have received great 
attention in 
theoretical computer science and statistics,
\cite{Va84,Va84a,Ri78,Ri86,Ri84,Ri87a,KV}.
On the other hand, %
the design of real learning systems seems to be dominated by
\it ad hoc %
\rm trial-and-error methods.
Applications of these recent 
theoretical results 
to real world learning system design are scarce and far between.
One exception is the elegant
paper by Quinlan and Rivest, \cite{QuRi89}.

In a companion paper \cite{ViLi99} we develop the theory and
mathematical validation of the MDL principle \cite{BRY98} based on ultimate
data compression up to the Kolmogorov complexity. 
Our purpose here is trying to bring theory and practice
together by testing the theory on simple real applications.
We give a brief accessible exposition of the theoretical background
of MDL and the mathematical proof that it works. The principle
is then applied to two distinct practical issues: that of
on-line character recognition and of robot modeling.
In both cases the issue is the supervised learning
of best model granularity for classification or extrapolation on
unseen examples. The systems and experiments are quite simple,
and are intended to just be demonstrations that the theoretical
approach works in practice. Both applications are in topical
areas and the results give evidence that the theoretical
approach can be extended to more demanding settings.

{\bf Contents:} From a theoretical point of view we explain
the general modeling principle of MDL and its relation
to Bayesianism. We show that
the MDL theory is  solidly based on a provably ideal method of
inference using Kolmogorov complexity and demonstrate that it is
valid under the assumption that the set of data is ``typical''
(in a precise rigorous sense) for the targeted model.
We then apply the theory to two experimental tasks that both concern
supervised learning of the best model granularity but are otherwise
quite dissimilar:
In the first task we want to obtain the best model sampling rate in
the design of an on-line hand-written character learning system,
and in the second task our goal is to determine
the best number of nodes in the hidden layer
of a three-layer feedforward neural network modeling a robot arm with
two degrees of freedom. It turns out that the
theoretically predicted optimal precision coincides with
the best experimentally determined precision. We conclude with
a discussion concerning the equivalence of
code-length based methods with probability-based methods, that is, 
of MDL and
Bayesianism.  

\subsection{Introduction to Learning On-Line Handwritten Characters}
One of the important aspects of AI research is machine 
cognition of various aspects
of natural human languages.
An enormous 
effort has been invested in the problem
of recognizing isolated handwritten characters or 
character strings, \cite{Su80}. Recognizing isolated hand-written characters
has applications, for example, in signature recognition and Chinese
character input. The alphanumerical
character learning experiment reported here 
is a pilot project which ultimately
aims at providing a practicable method to learn
Chinese characters. This problem knows
many practical difficulties---both quantitatively and qualitatively. 

There are several thousand independent
Chinese characters. No current key-board
input method is natural enough for casual users. Some of
these methods require
the user to memorize a separate code for each of seven thousand characters.
Some methods require the user to know %
{\em ping ying}---the phonological representation of
mandarin Chinese in latin characters.
\rm 
But then the translation into computer representation of characters
is not easy
because there are too many homonyms. Similarly, sound recognition
techniques do not help much either because 
almost every commonly used
Chinese character has more than one commonly used homonym. For
non-professional casual users,
hand-written input which is mechanically scanned and processed,
seems to be a quite reasonable way of entering data
into the computer.

A variety of approaches and algorithms have been proposed
to achieve a high recognition rate. The recognition process 
is usually divided into two steps: 

\begin{enumerate}
\item
feature extraction from the sample
characters, and 
\item
classification of unknown characters. 
\end{enumerate}

The latter
often uses either deterministic or statistical inference based on
the sample data, and various known mathematical and statistical
approaches can be used. Our contribution is on that level.
This leaves the technical problem of
feature extraction, whose purpose is to capture the essence from the
raw data. The state of the art is more like art than science.
Here we use existing methods which we explain now.

\subsubsection{Feature Extraction}
The most common significant features extracted
are center of gravity, 
moments, distribution of points, character loci, planar curve transformation,
coordinates, slopes and curvatures at certain points along the character
curve. The obvious difficulty of the
recognition task is the variability involved in handwritten characters. Not 
only does the shape of the characters depend on the writing style which
varies from person to person, but even for the same person trying to 
write consistently writing style changes with time.

One way to deal with this problem is the idea of `elastic matching'
\cite{Ku87,Ta82}.
Roughly speaking, the elastic matching method
takes the coordinates or slopes of certain points approximately
equally spaced along the curve of the character drawing as feature 
to establish the character
prototypes. To classify an unknown character drawing, the machine compares
the drawing with all the prototypes in its knowledge base according
to some distance function and the entered character is classified
as the character represented by the closest
prototype.
When an unknown character is compared to a prototype, 
the comparison of the features is not only made strictly
between the corresponding points with the prototype
but also with the 
points adjacent to the corresponding point in the prototype. 
The method we use for feature extraction is a new modification
of the standard elastic matching method.

\subsubsection{Classification}
Each implemented algorithm for character recognition embodies a
model or a family of models for character recognition. 
One problem of the extant research is
the lack of a common basis
for evaluation and comparison among various techniques. This is
especially true for on-line character recognition due to the lack of common 
standard and limited raw data source. A model from a family of
models induced by a particular method
of feature extraction is usually specified by a set of 
parameters. Varying the parameters gives a class of models with
similar characteristics. 

Consider the above mentioned elastic matching.
It uses certain points along the character curve as features. The interval size
used to extract these points along the curve is a parameter. How to determine
the value of this parameter which gives optimal recognition? 

Practically speaking, we can set the interval size to different
values and experiment on a given sample set of data to 
see which value gives the best performance. However, since the experiment 
is based on one particular set of data, we do not know if this interval size
value gives a similar optimal performance for all possible observations from
the same data source. A theory is needed to guide the parameter selection 
in order to obtain the best model from the given class of models.

\subsubsection{Model Selection}
Suppose we have models $M_1 , M_2 , \ldots$.
Let $H_i$ be the hypothesis `$M_i$ gives the best recognition rate'.
Our problem in selecting the best model consists in finding
the most likely hypothesis. We use
the {\em Minimum Description Length} principle
(referred as MDL hereafter) for this
purpose.
MDL finds its root in the well-known Bayesian inference and 
not so well-known Kolmogorov complexity.

Below we give the classic ``Bayes's rule.''
According to Bayes's rule, a specific hypothesis is preferred
if the probability of that hypothesis takes maximum value for a
given set of data and a given prior probability distribution over
the set of hypotheses. This happens for the hypothesis under which the
product of
the conditional probability of 
the data for the given hypothesis and the prior probability 
of the hypothesis is maximal.
When we take  the negative
logarithm of Bayes's formula, then this maximal probability is achieved
by the hypothesis under which the sum of the following two terms
is minimized: 
the description length of the error of the data for the given hypothesis and 
the description length of the model (the hypothesis).
Therefore, finding a maximum
value of the conditional probability of a given set of 
hypotheses and data becomes minimizing the combined
complexity or description length of the error and the model for a given set
of candidate models.

To quantify this idea,
the two description lengths are expressed in terms of the coding
length of the model (set of prototypes)
and the coding length of the error 
(combined length of all data failed to be described by the model).
The trade-off between simplicity and complexity of both
quantities is as follows.
\begin{enumerate}
\item
If a model is too simple, in the sense of having too short an encoding,
it may fail to capture the 
essence of the mechanism generating the data, resulting
in increased error coding lengths. 
\item
If a model
is too complicated, in the sense of having too long code length
(like when it consists of a table of all data),
it may contain a lot of redundance from the data and become too sensitive
to minor irregularities to give accurate predictions of the future
data. 
\end{enumerate}

The MDL principle states that among the given set of models, the
one with the minimum combined description lengths of both the model
and the error for given set of data is the best approximation
of the mechanism behind data and can be used to predict
the future data with best accuracy.

The objective of this work is to implement a small system
which learns to recognize
handwritten alphanumericals based on both
elastic matching and statistical inference. 
MDL is used to guide the model selection, specifically
the selection of the interval of feature extraction.
The result is then tested experimentally 
to validate the application of the theory.

\subsection{Introduction to Modeling a Robot Arm}
We consider the problem of modeling a robot arm consisting
of two joints and two stiff limbs connected as: joint, limb, joint, limb.
The entire arm moves in a fixed two-dimensional plane.
The first joint is
fixed at the origin. The position
of the other end of the arm is determined by the lengths of the two limbs,
together with
the angle of rotation in the first joint of the first
limb, and the angle of rotation in the second joint of
the second limb with respect to the first limb. 
The mobile end of the arm
thus has two degrees of
freedom given by the two angles.
In \cite{mckay92b} this problem is modeled using a Bayesian framework
to obtain a backpropagation network model.
We use MDL to obtain the best number of nodes in
the hidden layer of a three layer feedforward network model.
The method is essentially the same as in the character recognition experiment.
Just as before it is validated on a test set on unseen data.
\section{Theoretic Preliminaries}
We first explain the idea of Bayesian reasoning and give
``ideal MDL'' as a  noncomputable but provably good approach to
learning. Then, we will dilute the approach to
obtain a feasible modification of it in the form
of the real MDL. 
For another viewpoint of the relation between Bayesianism and
MDL see \cite{Bi95}.
In the later sections we apply MDL to learning best model granularity.
\subsection{Bayesianism}

Bayesianism is an induction principle
with a faultless derivation, yet allows
us to estimate the relative likelihood of different
possible hypotheses---which is hard or impossible with the
commonly used Pearson-Neyman testing.
With the latter tests we accept or reject a zero hypothesis
with a given confidence. If we reject the zero hypothesis,
then this does not mean that we do accept the alternative hypothesis.
We cannot even use the same data to test the alternative hypthesis.
(Or a subhypothesis of the alternative hypothesis---because
note that all hypotheses
different from the zero hypothesis must be taken together to
form the alternative hypothesis.)
In fact, this type of testing does not establish the
relative likelihood between competing hypotheses at all.

\begin{definition}
\rm
Consider a discrete sample space $\Omega$. Let $D, H_1 , H_2 ,  \ldots $
be a countable set of events (subsets) of $\Omega$.
${\bf H} = \{ H_1 , H_2 ,  \ldots   \} $ is called
\it hypotheses space\index{hypotheses space}
\rm . The hypotheses $H_i$ are
exhaustive (at least one is true).
From the definition of conditional probability,
that is, $P(A|B)=P(A  \cap  B)/P(B)$, it is easy to
derive %
\bf Bayes's formula %
\rm (rewrite $P(A  \cap  B)$ in two different ways):
\begin{equation}\label{eq1}
P(H_i |D)= {P(D|H_i )P(H_i ) \over P( D)} .
\end{equation}
\end{definition}
\noindent
If the hypotheses are mutually exclusive
($H_i   \cap   H_j = \emptyset$ for all $i,j$), then
$$
P(D)= \sum_i P(D|H_i )P(H_i ).
$$
Despite the fact that Bayes's rule is just a rewriting of the definition of
conditional probability,
its interpretation and applications are most profound and have caused
bitter controversy over the past two centuries.
In Equation~\ref{eq1}, the
$H_i$'s represent the possible alternative hypotheses
concerning the phenomenon we wish to discover. The term $D$ represents
the empirically or otherwise known data concerning this phenomenon.
The term $P(D)$, the probability of data $D$,
may be considered as a normalizing factor so that $\sum_i P(H_i |D) = 1$.
The term $P(H_i )$ is called the %
\it a priori %
\rm probability\index{probability!prior}
or $initial$ probability of hypothesis $H_i$, that is, it is
the probability of $H_i$ being true before we see any data.
The term $P(H_i |D)$ is called {\it a posteriori} or
{\it inferred} probability\index{probability!a posteriori}
In {\em model selection} we want to select the
hypothesis (model) with the maximum a posteriori probability (MAP).
\footnote{If we want to {\em predict} then we determine the expected
 a posteriori probability by integrating over hypotheses 
rather than choosing one hypothesis which maximises the
posterior.}

The most interesting term is the prior
probability\index{probability!prior} $P(H_i )$.
In the context of machine learning, $P(H_i )$ is
often considered as the learner's %
\it initial degree of belief %
\rm
in hypothesis $H_i$.
In essence Bayes's rule is a %
\it mapping %
\rm from %
\it a priori %
\rm probability\index{probability!prior}
$P(H_i )$ to %
\it a posteriori %
\rm probability\index{probability!a posteriori}
$P(H_i |D)$
determined by data $D$. In general, the problem is not so much that
in the limit the inferred hypothesis would not concentrate on the
true hypothesis, but that the inferred probability gives as much information
as possible about the possible hypotheses from only a limited number of data.
In fact, the continuous acrimonious debate between the Bayesian and non-Bayesian
opinions centered on the prior probability\index{probability!prior}.
The controversy is caused by the fact that Bayesian theory does
not say how to initially derive the
prior probabilities\index{probability!prior} for the hypotheses.
Rather, Bayes's rule only tells how they are to be %
\it updated%
\rm .
In the real world problems, the prior proabilities\index{probability!prior}
may be unknown, uncomputable, or even conceivably non-existent.
(What is the prior probability\index{probability!prior} of use of
a word in written English?
There are many different sources of different social backgrounds
living in different ages.)
This problem would be solved if we can
find a %
\it single %
\rm probability distribution to use as
the prior distribution in each different case, with
approximately the same result as if we had used the real
distribution. Surprisingly, this turns out to be
possible up to some mild restrictions.

\subsection{Kolmogorov Complexity}
So as not to divert from the main
thrust of the paper, we recapitulate the basic formal
definitions and notations in Appendixes~\ref{app.A}, \ref{app.B}.
Here we give an informal overview.

{\bf Universal description length:}
For us,
descriptions are finite binary strings. 
Since we want to be able determine where a description ends,
we require that the set of descriptions is a {\em prefix code}:
no description is a proper initial segment (proper prefix) of
another description.
Intuitively, the 
{\em prefix Kolmogorov complexity} of a finite object $x$ conditional $y$
is the length $K(x|y)$ in bits of the shortest effective description of $x$
using $y$ as input.  Thus, for every fixed $y$ the set of such shortest
effective descriptions is required to be a prefix code.
We define $K(x)=K(x| \epsilon )$ where $\epsilon$
means ``zero input''. 
Shortest effective descriptions are ``effective''
in the sense that we can compute the described objects from them.
Unfortunately, \cite{Ko65},
there is no general method to compute the length of a shortest description
(the prefix Kolmogorov complexity) from the object
being described. This obviously impedes actual use. Instead, one
needs to consider computable approximations to shortest descriptions,
for example
by restricting the allowable approximation time. This course is
followed in one sense or another in the practical incarnations such as
MDL. There one often uses simply the Shannon-Fano code \cite{CT91,LiVibook},
which assigns prefix code length $l_x := - \log P(x)$ to $x$ irrespective
of the regularities in $x$. If $P(x)=2^{-l_x}$ for
every $x \in \{0,1\}^n$, then the code word length of an
all-zero $x$ equals the code word length of a truly
irregular $x$. While the Shannon-Fano code gives an expected
code word length close to the entropy, it does not distinguish
the regular elements of a probability ensemble from the random
ones, by compressing individual regular objects more than the irregular ones.
The prefix code consisting of shortest prefix-free
programs with the prefix Kolmogorov complexities
as the code word length set does both: for every computable
distribution $P$ the $P$-expected code-word length
(prefix Kolmogorov complexity) is close to the entropy of $P$
{\em as well} as that every individual element is compressed as much
as is possible using an effective code.

{\bf Universal probability distribution:}
Just as the Kolmogorov complexity measures the shortest effective
description length of an object, the {\em algorithmic universal 
probability} ${\bf m} (x|y)$
of $x$ conditional $y$
measures the greatest effective probability of $x$ conditional $y$. 
It turns out that we can set ${\bf m} (x|y) = 2^{-K(x|y)}$,
(\ref{eq.m}) in the Appendix~\ref{app.B}. For precise definitions
of the notion of ``greatest effective probability''  the reader
is referred to this appendix as wel. 
It expresses a property of the probability 
of every individual $x$, rather than entropy which measures
an ``average'' or ``expectation'' over the entire ensemble of elements
but does not tell what happens to the individual elements. 
\footnote{
As an aside, for every fixed conditional $y$ the 
entropy $- \sum_x {\bf m}(x|y) \log {\bf m}(x|y) = \infty$.
}
We will use the algorithmic universal probability
as a universal prior in Bayes's rule to analyze ideal MDL.

{\bf Individual randomness:}
The common meaning of a ``random object'' is an outcome
of a random source. Such outcomes have expected properties
but particular outcomes may or may not possess these expected
properties. In contrast, we use
the notion of {\em randomness of individual objects}. 
This elusive notion's long
history goes back to the initial attempts by von Mises,
\cite{Mi19}, to formulate
the principles of application of the calculus of probabilities to
real-world phenomena.
Classical probability theory
cannot even express the notion of ``randomness of individual objects.''
Following almost half a century of unsuccessful attempts,
the theory of Kolmogorov complexity, \cite{Ko65}, and Martin-L\"of tests
for randomness, \cite{ML66}, finally succeeded in formally
expressing the novel notion of
individual randomness in a correct manner, see \cite{LiVibook}.
 Every individually random object
possesses individually all effectively
testable properties that are only expected for outcomes of
the random source concerned. It is ``typical'' or ``in general
position'' in that it will
satisfy {\em all} effective tests for randomness---
known and unknown alike. A major result states that
an object $x$ is individually random
with respect to a conditional probability distribution $P(\cdot |y)$
iff $\log ({\bf m}(x|y)/P(x|y))$ is close to zero. In particular
this means that $x$ is ``typical'' or ``in general position''
with respect to conditional distribution $P( \cdot |y)$ iff
the real probability $P(x|y)$ is close to the
algorithmic universal probability ${\bf m}(x|y)= 2^{- K(x|y)}$.
That is, the prefix Kolmogorov complexity $K(x|y)$ is
close to the Shannon-Fano code length of $x$ as element
of the a set with probability distribution $P( \cdot |y)$.

For example, if $H$
is the hypothesis that we deal with a fair coin and the data sample
$D$ is a hundred outcomes `heads' in a row, then $D$ isn't typical
for $H$. But if $D$ is a truly random individual sequence with respect
to $H$ (a notion that has a precise formal and quantitative meaning),
then $D$ is typical for $H$. The probability of atypical sequences
is very small and goes to zero when the data sample grows unboundedly.

{\bf Prediction and Model Selection:}
It has been shown by Solomonoff \cite{So78} that the
continuous variant of ${\bf m}$
has astonishing performance in predicting sequences where the
probability of the next element is computable from the initial
segment.
We now come to the punch line: Bayes's rule using
the algorithmic universal prior distribution,  suggested
by Solomonoff already in \cite{So64}, yields
Occam's Razor\index{Occam's Razor principle} principle and
is rigorously shown to work correctly in the companion paper
\cite{ViLi99}. Namely,
it is shown that this implies that data compression is almost always
the best strategy, both in hypothesis identification and prediction.

\subsection{Minimum Description Length Principle}
Scientists formulate their theories in two steps. Firstly, a scientist,
based on scientific observations,
formulates alternative hypotheses (there can be an infinity
of alternatives), and
secondly a definite hypothesis is selected. The second
step is the subject of
inference in statistics. 
Historically this was done by many different principles, like
Fisher's Maximum Likelihood
principle, various ways of using
Bayesian formula (with different prior distributions). Among the most
dominant ones is the `common sense' idea
of applying Occam's razor principle of
choosing the simplest consistent theory.
But what is ``simple''? We equate ``simplicity'' with ``shortness
of binary description,'' thus reducing the razor to objective 
data compression.

However, no single principle is both theoretically sound and
practically satisfiable in all situations.
Fisher's principle ignores the prior probability distribution (of hypotheses).
To apply Bayes's rule is difficult because we usually 
do not know the actual
prior probability distribution.
(What is the prior distribution of words in written English,
where there are many sources of many ages and social classes?)
No single principle turned out to be satisfiable in all situations.
Philosophically speaking,
relative shortness achievable by ultimate data compression
 presents an ideal way of
solving induction problems. 
However, due to the non-computability of the Kolmogorov complexity
and the associated algorithmic universal prior
function, such a theory cannot be directly used.
Some approximation is needed in the real world applications.

Rissanen~\cite{Ri78}
follows Solomonoff\index{Solomonoff, R.J.}'s idea, but substitutes a
`good' computable approximation
to ${\bf m} (x)$ to obtain the so-called
{\bf Minimum Description Length}
\rm principle.
Rissanen not only gives the principle, more importantly he also gives
the detailed formulas on how to use
this principle. This made it possible to use the MDL principle.
The basic form of the MDL principle can be intuitively stated as follows:

\begin{quotation}
{\bf Minimum Description Length Principle.}
{\it The best theory to explain a set of data is the one which minimizes
the sum of}\\
$\bullet$ {\it the length, in bits, of the description of the theory;}\\
$\bullet$ {\it the length, in bits, of data when encoded with the help of the theory.}
\end{quotation}

A survey of the development of the MDL principle in statistical
inference and its applications is given in \cite{BRY98}.
In \cite{ViLi99} 
the relationship between the Bayesian approach
and the minimum description length approach is established.
The general modeling principle
MDL is sharpened and clarified, 
abstracted as the ideal MDL principle and defined from Bayes's
rule by means of Kolmogorov complexity. 
The argument runs as follows:

Given a data sample and a family of models (hypotheses) one wants to select
the model that produced the data. A priori it is possible that
the data is atypical for the model that actually produced it.
Meaningful induction is possible only by ignoring this possibility.
Strictly speaking, selection
of a ``true'' model is improper usage,
``modeling the data'' irrespective of truth and falsehood of
the resulting model is more appropriate. In fact, given data sample
and model class the truth about the models is impossible to ascertain
and modeling as well as possible is all we can hope for.
Thus, one wants to select a model 
for which the data is
typical. 
The best models make the two-part description
of the data using the model as concise as possible.
The simplest one is best in accordance
with Occam's razor principle since it summarizes the relevant
properties of the data as concisely as possible. In probabilistic data or data
subject to noise this involves separating regularities (structure)
in the data from random effects. 

From Bayes's Formula~\ref{eq1}, we must choose the hypothesis $H$
that maximizes the posterior $P(H|D)$.
Taking the negative logarithm
on both sides of Equation~\ref{eq1}:
$$
- \log  P(H|D) = - \log  P(D|H) - \log  P(H) + \log  P(D) .
$$
Here, $\log P(D)$ is a constant and can be ignored because we just
want to optimize the left-hand side of the equation over $H$.
Maximizing the $P(H|D)$'s over all possible $H$'s is equivalent to
{\it minimizing} $- \log  P(H|D)$, that is, minimizing
$$
- \log  P(D|H) - \log  P(H).
$$
To obtain the ideal MDL principle it suffices to replace the terms in the sum by
$K(D|H)$ and $K(H)$, respectively. In view of (\ref{eq.m}) in the 
Appendix~\ref{app.B}
this is justified provided $- \log P(D|H) \eqa - \log {\bf m}(D|H)$
and also $- \log P(H) \eqa - \log {\bf m}(H)$. In \cite{ViLi99} we show that 
the basic condition
under which this substitution is justified is encapsulated
as the Fundamental Inequality, which in broad terms states that the
substitution is valid when the data are random, relative to every
contemplated hypothesis and also these hypotheses are random relative to
the (universal) prior. Basically, the ideal MDL principle states
that the prior probability
associated with the model should be given by the algorithmic universal
probability, and the sum of the log universal probability
of the model plus the log of the  probability of the data given the model
should be minimized. For technical reasons the latter
probability $P(D|H)$ must be computable. 

It is important to note that using the algorithmic universal prior
we compress every model $H$ to its prefix
Kolmogorov complexity $K(H) = - \log {\bf m}(H)$.
Applying the ideal MDL principle then compresses de description of
the data encoded using the model, $D|H$, to its prefix
Kolmogorov complexity $K(D|H) = - \log {\bf m}(D|H)$ as well
for the model $H$ minimizing the sum of the two complexities. 
Roughly speaking, the MDL selection assumes that the data set
is ``typical'' for the selected model $H$. Thus, MDL aims
at selecting a model for which the data
are ``typical'', even if there happened to be
a different ``true'' model that inappropriately
generated ``atypical'' data. In this manner application of MDL
is resilient to overfitting the model.

\subsection{Ideal MDL versus Real MDL}
Using the algorithmic universal
prior, the ideal MDL principle is valid for a set of data samples
of Lebesgue measure one, the ``random'', ``typical'' outcomes,
for every contemplated hypothesis. For these ``typical'' outcomes
we have $K(D|H) \eqa - \log P(D|H)$ which means that the classic
Shannon-Fano code length reaches the prefix Kolmogorov
complexity on these data samples. The Shannon-Fano code 
that assigns code words of length $\eqa - \log P(\cdot)$
to elements randomly drawn according to a probability density $P(\cdot)$
is in fact used in the applied statistical version of MDL.
Thus, under the assumption that the data sample is typical for the contemplated
hypotheses, the ideal MDL principle and the applied statistical one coincide,
and moreover, both are valid for a set of data samples of 
Lebesgue measure one \cite{ViLi99}.
The latter result has been obtained in the statistical theory 
using probabilistic arguments \cite{BRY98}.

The term $- \log  P(D|H)$, also known as the %
\it self-information %
\rm in information theory and the negative log likelihood in statistics,
can now be regarded as the number of bits it takes to redescribe
or encode $D$ with an ideal code relative to $H$.
In different applications, the
hypothesis $H$ can mean many different things, such as
decision trees, finite automata, Boolean formulas, or a
polynomial.
\begin{example}
\rm
In general statistical applications, one assumes that $H$ is some model
$H( \theta )$ with a set of parameters
$\theta =  \{  \theta_1 ,  \ldots  , \theta_k  \} $ of precision $c$,
where the number $k$ may vary and influence the descriptional complexity
of $H( \theta )$. For example, if we want to determine the distribution
of the length of beans, then $H$ is a normal distribution $N( \mu , \sigma )$
with parameters median $\mu$ and variation $\sigma$.
So essentially we have to determine the correct hypothesis
described by identifying the type
of distribution (normal) and the correct parameter vector $( \mu , \sigma )$.

In such cases, we minimize
$$
- \log  P(D | \theta ) - \log  P( \theta ).
$$

\end{example}
\begin{example}
\rm
Let's consider the fitting of a `best' polynomial
on $n$ given sample points in the 2-dimensional plane. 
This question is notoriously underdefined, since both a 1st degree
polynomial with $\chi^2$ best fit, and a $(n-1)$th degree polynomial
with perfect fit are arguably the right solutions.
But with the MDL principle we can find an objective `best'
polynomial among the polynomials of all degrees.

For each fixed $k$, $k=0,  \ldots  , n-1$,
let $f_k$ be the best polynomial of degree $k$,
fitted on points $(x_i ,y_i )$ ($1 \leq i \leq n$),
which minimizes the error
$$
error(f_k ) = \sum_{i=1}^n (f_k (x_i ) - y_i )^2 .
$$
Assume each coefficient takes $c$ bits. So $f_k$ is encoded in
$c(k+1)$ bits. Let us interpret the $y_i$'s as
measurements for argument $x_i$ of some true polynomial
to be determined. Assume that the  measurement process
involves errors. Such errors are accounted for by
the commonly used Gaussian (normal)
distribution\index{distribution!normal} of the
error on $y_i$'s. Thus, given that $f$ is the true polynomial,
$$
\Pr (y_1 ,  \ldots  , y_n | f, x_1, \ldots ,x_n )
= \prod \exp (- O((f (x_i ) - y_i )^2 ) ) .
$$
The negative logarithm of above is $c' \cdot error(f )$ for some
computable $c'$. The MDL principle tells us to choose $f=f_m$, with
$m  \in   \{ 0,  \ldots  ,n-1  \} $, which
minimizes $c (m+1) + c' \cdot error(f_m )$.
\end{example}

In the original Solomonoff approach a hypothesis $H$ is
a Turing machine. In general we must avoid such a too
general approach in order to keep things computable. In different
applications, $H$ can mean many different things.
For example, if we infer decision trees, then $H$ is
a decision tree. In case
of learning Boolean formulas, then $H$ may be a Boolean formula.
If we are fitting a polynomial curve to a set of data, then $H$ may be a 
polynomial of some degree. In Experiment 1 below, $H$ will be the model for
a particular character.
Each such $H$ can be encoded by a binary string from a prefix-free set,
where a set of codes is %
 prefix-free %
\rm if no code in the set is a prefix
of another. 
\section{Experiment 1: On-Line Handwritten Characters}
\subsection{Model Development}
\subsubsection{Basic Assumptions}
When an alphanumeral character is drawn on a planar surface,
it can be viewed as a composite planar curve, the shape of which is
completely determined by the coordinates of the sequence of 
points along the curve. The order of the sequence is determined
by on-line processing the data from the scanning
machinery at the time of writing the character.
Since the shape tends to vary from person to person and 
from time to time, so do the coordinates of the point 
sequence. A key assumption in our treatment is
that for a particular person writing consistently the shape
of the curve tends to converge to an average shape, in the sense that
the means of corresponding coordinates of 
the sampled point sequences converge.

That is, we assume:
\begin{itemize}
\item
Each shape of an example curve for
a particular character contains a set
distinguished feature points. 
\item
For each such point,
the average of the instances in the different examples
converges to
a mean. 
\item
Moreover, there is a {\em fixed} probability 
distribution (possibly unknown) for each such point
which is symmetric about
the mean value, and the variance is assumed to
be the same for all the character drawings. 
\end{itemize}
Essentially, we only assume
that 
one person's hand-writing has a
\it fixed %
\rm associated probability distribution, which
does not change.
\subsubsection{Feature Space, Feature Extraction and Prototypes}
A Kurta ISONE digitizer tablet with 200/inch resolution
in both horizontal and vertical directions was used
as the scanner to obtain and send the coordinates of the sequential points
of the character curve
on the tablet to the microprocessor of a IBM PS/2 model 30
computer. The system was implemented using programming language C.
The coordinates were normalized on a
30 by 30 grid in horizontal and vertical directions. 

The character drawn on the tablet is processed on-line.
The sequence of the coordinates in order of time
of entry is stored in the form of a linked
list. This list is preprocessed in order to 
remove the repeating points due to hesitation at
the time of writing, and to fill in the gaps between
sampled points resulted from the sampling rate limit
of the tablet. 

The latter needs some explanation: the 
digitizer has a maximum sampling rate of 
100 points/second. If a person writes a character
in 0.2 seconds, only 20 points on the character curve will be
sampled, leaving gaps between those points.

The preprocessing procedure ensures that
in the resulting linked list
any pair of consecutive points on the curve
has at least one component
of coordinates (stored as integers between 0 and 30)
differing by 1 and no coordinate
component differing by more than 1. For a preprocessed list
$((x_1 ,y_1) , \ldots , (x_n ,y_n) )$ therefore we have 
that for all $i$ ($1 \leq i < n$)
\begin{eqnarray*}
|x_i - x_{i+1} | + |y_i - y_{i+1}| & \geq & 1 \\
|x_i - x_{i+1} | , |y_i - y_{i+1}| & \leq & 1 
\end{eqnarray*}
The preprocessed
curve coordinate list is then sent to the 
feature extraction process. So
far the coordinates are still integers in the range of
0 to 30.

The coordinates of certain points along the character 
curves are taken as relevant {\em features}. Feature extraction
is done as follows. A character may consist of more 
than one stroke (a stroke is the trace from a pen 
drop-down to pen lift-up), the starting and  
ending  points of every stroke are mandatorily taken as features.
In between, feature points are taken at a fixed
interval, say, one point for every $n$ points
along the preprocessed curve, where $n$ is called
\bf feature extraction interval\rm%
. This is to 
ensure that the feature points are roughly equally
spaced. Actually the Euclidean length between any two
points on the stroke curve, 
excluding the last point of a stroke, varies from 
$n$ to $\sqrt 2 n$ (for the diagonal). 

The sequence of the feature point coordinates extracted from a given
character drawing constitute
a {\em feature vector}. (If the character drawing contains more than
one stroke, its feature vector consists of the concatenation
of the feature vectors of the individual strokes in time-order.)
The dimension of the feature vector is the number of entries in it---or
rather twice that number since each entry has two coordinate components. 
Obviously the dimension of the feature vector is
also a random variable since the shape and the total
number of points on the character curve varies
from time to time. The dimension of the feature vector
is largely determined by the feature extraction interval.

The extracted feature vector of a character is viewed
as a {\em prototype} of a character, and is stored 
in the knowledge base of
the system as such.

\subsubsection{Comparison between Feature Vectors}
Before the system is employed to recognize characters, it
must first learn them. It is
trained with examples of the character drawings from the same data source
which it is supposed to recognize afterwards. 
Here the `same data source' means
the same person writing consistently.
The basic technique used in both training and recognition is
the comparison or matching between prototypes or feature vectors. 
To compare any two prototypes or feature vectors of equal dimension, 
we can simply
take the Euclidean distance 
between the two vectors. Mathematically 
this means 
subtracting each component of one vector from its corresponding 
component in the other feature vector, summing up the square of the 
differences and taking the square root of the sum. If the two
prototypes are $\chi = ((x_1 ,y_1) \ldots , (x_n ,y_n))$ and 
$\chi' = ((x_1' ,y_1') \ldots , (x_n' ,y_n'))$, then the
distance between them is
$$
\sqrt { \sum_{i=1}^n (x_i - x_i' )^2 + (y_i - y'_i )^2 } .
$$
The knowledge base of the system is a collection of 
feature vectors stored in the form of a linked list. Each
such feature vector is an example of a
particular character and is called a {\em prototype} for that
character. Each newly entered character drawing in
the form of a feature vector is compared to the prototypes
in the knowledge base.
But we do not (cannot)
assume that all feature vectors
extracted from examples 
of the same character will have the same dimension.

Therefore, the comparison 
technique used in our system follows 
the spirit of this mathematical definition
but is more
elastic.
As a consequence, corresponding feature points may be
located in different places in the feature vectors. 
This problem is solved by so-called
{\em elastic matching} which compares a newly sampled
feature vector with the set of stored feature vectors,
the {\em prototypes}. 
The elasticity is reflected in two aspects:
\begin{description}
\item[Dimension tolerance]
is a constant integer $T_d$ such that the new feature vector
is compared with all stored feature vector of which the dimension is not
more than 
$T_d$ different.
That is, if the new 
feature vector has $n$ feature points, it will be compared (matched) with
all the prototypes with a number of feature points 
in the range of $[n - T_d,  n + T_d ]$.  
\item[Local extensibility]
is an integer constant $N_e$
such that the $i$th feature point of the new feature vector
is compared with the feature points 
with index ranging from $i - N_e $
to $i + N_e $
of each prototype satisfying
the dimension tolerance.
The least Euclidean distance found this way is considered to be the 
`true' difference $d_i$ between the two vectors at $i$th feature point.
\end{description} 
\begin{definition}
\rm
If the dimension of the new feature vector $x$ is $n$, then the
{\em elastic distance} $\delta(x,x')$ between $x$ and a
prototype $x'$ is defined as
\[ \delta (x,x') = \sqrt{ \sum_{i=1}^n d_i^2 } \]
if $x'$ is within the dimension tolerance $T_d$ of $x$,
and  $\delta (x,x') = \infty$ otherwise.
\end{definition}

For our particular problem, experimental evidence
indicates that it suffices to set both $T_d$ and $N_e$ to 1.
In our experiment we used elastic distance between a new feature vector
$x$ and a prototype $x'$ as
computed above with
$T_d = N_e =1$.

\subsubsection{Knowledge Base and Learning}
The knowledge base is constructed in the learning
phase of the system
by sampling feature vectors of
handwritten characters while telling the system which character
it is an example of. Our system
uses the following {\bf Learning Algorithm} to establish the knowledge base.

\begin{description}
\item[Step 0.]
Initialize the knowledge base $S$ to the empty set $\emptyset$.
(The elements of $S$ will be triples $(x, \chi, c )$ with $x$ a
preprocessed feature vector (a prototype), $\chi$ is the character
value of which $x$ is a prototype, and $c$ is a counter.)
Assign values to weights $\alpha, \beta$ (used later to combine prototypes)
so that $\alpha + \beta = 1$.
\item[Step 1.]
Sample a new example
of a character feature vector, say $x$ after preprocessing, together
with its character value, say $\chi$. (Actually, the user draws a new example
handwritten character on the tablet and indicates the character
value---which character the drawing represents---to the system.)
\item[Step 2.]
Check $S$ whether or not
any prototypes exist for character $\chi$.

If there is no prototype for $\chi$ in $S$, then
store $x$ in $S$ as a prototype for $\chi$ by setting
\[ S := S \cup \{(x, \chi , 1)\}. \]

If $P_x = \{y , \ldots ,z \}$ 
is a nonempty list of prototypes for $\chi$ in $S$,
then determine elastic distances $\delta (x,y) \ldots \delta (x,z)$.
Let $P_x^{min} \subseteq P_x$ be the set of prototypes in $P_x$ such that
for all $x' \in P_x^{min}$ we have
\[
\delta (x,x') = \delta_{min} \stackrel{\rm def}{=} 
\min \{ \delta (x,y): y \in P_x \}.
\]
Now $x' \in P_x^{min}$ may or may not be one of the prototypes with the 
character value $\chi$. 

\subitem
{\bf Step 2.1.}
If $x' \in P_x^{min}$ and $(x', \chi ,m) \in S$
(the minimum distance $\delta_{min}$
is between $x$ and one of
the prototypes for $\chi$ in $S$), then
\[ x' := \alpha x' + \beta x; m := m+1 . \]
(The new prototype is combined with the existing
prototype by taking the weighted average of every coordinate
to produce a modified prototype for that character. Moreover,
we add one to the counter associated with this prototype)
\subitem
{\bf Step 2.2.}
If Step 2.1 is not applicable,
then for no $x' \in P_x^{min}$ we have $(x', \chi, \cdot ) \in S$
(the minimum distance is between $x$ and 
prototype(s) in $S$ with character
value $\neq \chi$). Then we distinguish two cases.

{\em Case 1.} There is $(x' , \chi ,m) \in S$
such that 
$\delta (x,x') \leq \delta_{min} (m + 1)/m $.
(Here $m$ is number of character drawings
which have consecutively be combined to form the current $x'$ prototype.)
Then set
\[ x' := \alpha x' + \beta x; m := m+1 . \]
It is expected that
the modified prototype will have minimal distance
$\delta_{min}$ next time 
when a similar drawing of the same character value arrives.

{\em Case 2.}
The condition in Case 1 is not satisfied. Then
the new prototype will be saved in the knowledge base
as a new prototype for the character by setting
\[ S := S \cup \{(x, \chi , 1)\}. \]
\end{description}
Notice that more than one prototype for a single character
is allowed in the knowledge base.
\subsubsection{Recognition of an Unknown Character Drawing}
The recognition algorithm is simple. Assume the Learning Algorithm
above has been executed, and a knowledge base $S$ with
prototypes of all possible characters has been constructed.
When an new character drawing is presented to the system, it
is compared to all the prototypes in the knowledge base
with dimension variation within
the range specified by the dimension tolerance. 
The character of the prototype which has minimum
$\delta$-distance from the presented feature vector
is considered to be the character value
of that feature vector. The rationale here
is that the prototypes are considered as the `mean' values of the 
feature vectors of the characters, and the variances of 
the distribution are assumed to be the same for all prototypes.
Formally, the {\bf Recognition Algorithm} is as follows.
\begin{description}
\item[Step 0.]
Sample a new example
of a character feature vector, say $x$ after preprocessing.
(Actually, the user draws a new example
handwritten character on the tablet which is preprocessed to
form feature vector $x$.)
\item[Step 1.]
If $S = \{(x_1 , \chi_1, \cdot ) , \ldots ,(x_n , \chi_n, \cdot ) \}$
is the knowledge base,
then determine elastic distances $\delta (x,x_1) \ldots \delta (x,x_n)$.
If $\delta (x,x_i)$ is the minimal distance in this set, with $i$ 
is least in case more than one prototype induces minimum distance,
then set $\chi_i$ is the character value for $x$ and
\[\mbox{Recognize character } \chi_i . \]
\end{description}

This concludes the main procedure of training and classification.
A few remarks are in order to explain differences with the
original elastic matching method.

\begin{enumerate}
\item
This process differs from the original elastic matching method
in the the way of prototype construction. More than one prototype
are allowed for a single character. By our procedure
in the Learning Algorithm, a prototype is the {\em statistical
mean} of a number of positive examples of the character.
\item
Every prototype is a feature vector which in turn is a point
in the feature space of its dimension. Since the classification
is based on statistical inference, the rate of correct 
classification depends not only on how well the prototypes
in the knowledge base are constructed, but also on the variability
of the handwriting of the subject. Even though more than one
prototype is allowed for any character in the knowledge base,
too many prototypes may result in an overly dense feature space.
When the $\delta$-distance between two points (two prototypes
in the knowledge base) in the feature space is comparable to the
variability of the subjects handwriting, 
then the rate of correct classification
may drop considerably.
\item
The prototypes in the knowledge base constitute the model 
for the system. How well the prototypes are constructed will
essentially determine the rate of correct classification and
therefore the performance of the model. For the scheme described
above, the prototypes are constructed by extracting points at 
a constant interval. Generally speaking, more points in the
prototypes gives a more detailed image of the character drawing
but may also insert random `noise' in the model.
\par\noindent
Application of  MDL to guide the selection of `best'
feature extraction interval is the main thrust of
this work, to which we proceed now.
\end{enumerate}
\subsection{Implemented Description Lengths and Minimization}
\par\noindent
The expression in MDL consists of two terms: the model
and error coding lengths. The coding efficiency
for both of these two terms must be comparable, otherwise
minimizing the resulted expression of total description length
will give either too complicated or too simple models. For this
particular problem, the coding lengths are determined
by practical programming considerations. 

A set of 186 character drawings, 
exactly 3 for each of the 62 alphanumeral characters,
were processed to feature vectors and presented
to the Learning Algorithm, to form the raw database. 
The character drawings were stored in standardized integer
coordinate system standardized from 0 to 30 in both $x$ and $y$ axis.
After preprocessing as above,
they were then input to the Learning Algorithm
to establish a knowledge base: the collection of 
prototypes with normalized real coordinates,
based on a selected feature extraction
interval. 

Subsequent to the construction of the knowledge base,
the system was tested by having it classify the same set of character
drawings using the Recognition Algorithm. 
This procedure served to establish the error code length 
and the model code length
which are defined as follows.

\begin{definition}
\rm
The {\em error code length} or 
{\em exception complexity} is the sum of the total
number of points for all the incorrectly classified character
drawings. This represents the description of the data given
the hypothesis.

The {\em model code length} or {\em model complexity}
is the total number of points in all the prototypes in the 
machine's knowledge base multiplied by 2. 
This represents the hypothesis.

The {\em total code length} is the sum of the error code length
and the model code length.
\end{definition}

\begin{remark}
\rm
The factor of 2 in the model code length
is due to the fact that the prototype coordinates
are stored as real numbers which takes twice as much
memory (in programming language C) 
as the character drawing coordinates which are represented in
integer form. One might wonder why the prototype coordinates are 
real instead of integer numbers. The reason is to facilitate
the elastic matching to give small resolution for comparisons
of classification. 
\end{remark}

Thus, both the model and error code lengths
are directly
related to the feature extraction interval. The smaller 
this interval, the more complex the model, but the smaller
the error code length. The effect is reversed if 
the feature extraction interval goes toward larger values. Since
the total code length is the sum of 
the two code lengths, there should be a value of feature
extraction interval which minimizes the total code length.

This feature extraction interval is considered to be the `best' one
in the spirit of MDL. The corresponding model, the 
knowledge base, is considered to be optimal in the sense
that it contains enough essence from the raw data but
eliminates most redundancy due to noise from the 
raw data. This optimal feature extraction interval can
be experimentally determined by carrying out the 
above described build-and-test
(building the knowledge base and then test it based on the 
same set of characters on which it was built) for 
a number of different feature extraction intervals.

\begin{figure}
\hfill\ \psfig{figure=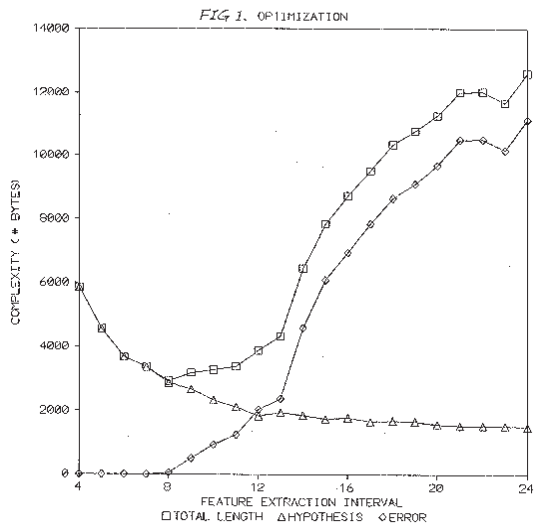,width=3in} \hfill\
\caption{Experimentally Determined Error- and Model Code Lengths}
\label{fig.1}
\end{figure}

The actual optimization process was implemented on the 
actual system we constructed, and available to the user.
For our particular set of characters and trial, the results
of classifying by the Recognition Algorithm
the same set of 186 character drawings used
by the Learning Algorithm to establish the knowledge base,
is given in Figure~\ref{fig.1}. 
Three quantities are depicted: 
the model code length, the error code length, and the total
code length, versus different feature extraction intervals (FEATURE EXTRACTION
INTERVAL in the figure). 
For larger feature extraction intervals, 
the model complexity is small but most of the character
drawings are misclassified, giving the very large 
error code length and hence the very large
\begin{figure}[htb]
\hfill\ \psfig{figure=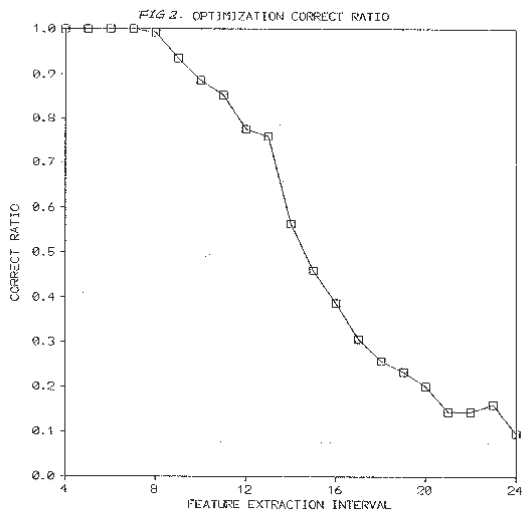,width=3in} \hfill\
\caption{Fraction Correctly Classified Training Data}
\label{fig.2}
\end{figure}
total code length. On the other hand, when the feature 
extraction interval is at its low extremal value, all
training characters get correctly classified which gives
zero error coding length. But now the model
complexity reaches its largest value, resulting also in
a large total code length again. The minimum total
code length occurred in our experiment at
an extraction interval of 8, which
gives 98.2 percent correct classification. Figure~\ref{fig.2}
illustrates the fraction of correctly classified character drawings for 
the training data.
\begin{figure}[htb]
\hfill\ \psfig{figure=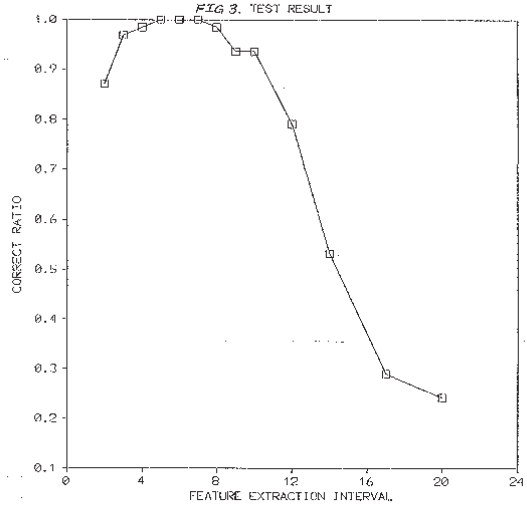,width=3in} \hfill\
\caption{Fraction Correctly Classified New Test Data}
\label{fig.3}
\end{figure}
\subsection{Validation of the Model}
\par\noindent
Whether the `optimal' model, determined by choosing the interval
yielding minimal total code length for the training data,  really
performs better than models in the same
class using different feature extraction intervals,
can be tested by classification of new data---new
character drawings. 

We have executed such a test by having
the set of 62 characters drawn anew
by the same person who provided the raw data base
to build the knowledge base. After preprocessing,
the feature vectors resulting from these data were
entered in the Recognition Algorithm. The new data
are considered to be from the same source as the previous
data set. 

This
new data set was classified by the system
using the knowledge bases built by the Learning Algorithm from the 
training data set of 186 character drawings, 
based on different feature extraction 
intervals. The test results are plotted in
Figure~\ref{fig.3} in terms of the fraction of 
correct classification (CORRECT RATIO) versus
feature extraction interval (FEATURE EXTRACTION INTERVAL)
It is interesting to see that a 100\% correct
classification occurred at feature extraction intervals
5, 6 and 7. These values of feature extraction intervals
are close to the optimal value 8 resulting from MDL considerations. 
Furthermore,
at the lower feature extraction intervals, 
the correct classification rate drops, indicating
the disturbance caused by too much redundance in
the model. The recommended working feature
extraction interval is thus either 7 or 8 for this 
particular type of character drawings.

\section{Experiment 2: Modeling a Robot Arm}
\label{sect.robot}
In the second experment,
the problem is to model a two-jointed robot arm described in the 
introduction. A mathematical description is as follows.
Let $r_1$ and $r_2$ be the lengths of the two limbs constituting
the arm. One end
of the limb of length $r_1$ is located in the joint
at the origin $(0,0)$ 
of the two-dimensional
plane in which the arm moves. The angle the limb makes
with the horizontal axis is $\theta_1$. The angle the limb of length
$r_2$, the second limb, makes with the first limb (in the second joint) is $\theta_2$.  Then the  relationship
between the coordinates $(y_1,y_2)$ of the free end
of the second limb (the hand so to speak) and the
variables $\theta_1, \theta_2$ is given by
\begin{eqnarray*}
y_1&  = & r_1 \cos (\theta_1)+r_2 \cos(\theta_{1}+\theta_{2}) \\
y_2 & = &
r_1 \sin (\theta_1)+r_2 \sin(\theta_1+\theta_2).
\end{eqnarray*}
The goal is to construct a feedforward neural network that correctly
associates the $(y_1 , y_2 )$ coordinates to the $(\theta_1 , \theta_2 )$
coordinates. 
\noindent

As in \cite{mckay92b} we set $r_1 = 2$ and $r_2=1.3$.
The setup is similar to the character recognition experiment
except that the data are not real-world but computer generated.
We generated random
examples of the relation between $y_1,y_2$
and $\theta_1 , \theta_2$ as in the above formula
and gaussian noise of magnitude 0.05 was added to the outputs.

Since we want the learned model to
extrapolate from the training examples rather than
interpolate between them, the
training sets consist of random examples
 taken from two limited and separate areas of
the domain. In the training data
the first angle $\theta_1$ was in between 90 and 150 degrees or
between 180 and 240 degrees, and the second angle was
in between 30 and 150 degrees.
To test 
extrapolation capabality of the learned model we used 
a unseen test set in which $\theta_1$ ranges between 0 and 270
and $\theta_2$ between 0 and 180 degrees.


\subsection{Model Features}
The model class consists of three-layer feedforward networks.
The first layer is the input layer consisting
of two input nodes with as input the real values of the two
angles $\theta_1 , \theta_2$. Both nodes in the input layer
are connected with every node in the second layer---the hidden layer---of
which the number of nodes is to be determined. Every node in the second
layer is connected to both nodes in the third (output) layer,
yielding the two real-valued output values $y_1, y_2$.
There are no other connections between pairs of nodes.
The second layer nodes have sigmoidial transfer functions
and in the third layer output nodes have linear transfer functions.
Thus, the only unknowns in the network are
the number of nodes in the hidden layer, the weights on the connections
and the biases of the nodes in the hidden layer.
For every number of $k$ nodes ($k=2,3, \ldots , 15$) in the hidden layer
we learned the weights and biases of the network using standard
methods by repeatingly presenting the training set.
After that, the learned models are evaluated expermentally as to their
prediction errors on the unseen test set.

During the experiments we noticed that if we used a test set
from the same domain as the training set---thus testing interpolation
rather than extrapolation---then
the increase of error
with increasing number of nodes in the hidden
layer (after the optimal number) was small.
For the unseen test set described earlier---testing extrapolation or
generalization---the increase of the error after the optimal
network size was more steep. 
Below we used the latter ``generalization'' test set.

\subsection{Determining Size of Hidden Layer by MDL}
We verify the contention that in this experimental setting
the hypothesis selected by the MDL principle using the training
data set can be expected to be a good
predictor for the classification of unseen data from a test set.

Neural networks can be coded in the following way.
Both the topology, the biases of the nodes, 
 and the weights on the links are coded.
Assume that the network
contains $k$ nodes.
The code starts with the number $k$. Next a list of $k$ bias values
is encoded using $l$ bits for each bias value. We need
$k \times (k-1)$ bits to describe which pairs of
nodes are connected by directed arcs (possibly in two ways).
The weight for each link is given using a
precision of $l$ bits. Concatenating all these
descriptions in a binary string we can only retrieve
the network if we can parse the constituent parts.
Keeping the above order of the constituents we can do
that if we know $k$. Therefore, we start the encoding
with a prefix-free code for $k$ in $\log k +
2 \log \log k$ bits.\footnote{This is standard in prefix-free
coding, see Appendix~\ref{app.A}.
}
The total description now takes at most
$\log k  + 2\log\log k  + k\times l + k(k-1) + m\times l$
bits, where $m$ is the number of directed edges (links).

For three-layer feedforward  networks that constitute our models,
with two input nodes and two output nodes
and $k$ nodes 
in the hidden layer, the topology is fixed. As already stated,
we have to choose only 
the weights on the links and the biases of the hidden nodes.
This gives descriptions
of length
$\log k  + 2\log\log k  +  5k \times l$ bits.
For the range of $k,l$ we consider the logarithmic terms
can be ignored. Thus, the model cost is set at $5kl$ bits,
and with precision $l=16$ the model cost is linear
in $k$ at $80k$ bits.

The encoding of the output data 
for the neural network, to determine the error cost in the
MDL setting, depends on whether
they are given as integers or reals. For integers one takes the 16 bits that
the MaxInt format requires, and for real numbers usually twice as many,
that is, 32 bits.
For reals such an encoding introduces a new problem: when is
the output correct? We consider it correct
if the real distance between the output vector and the target vector is
under a small fixed real value.  In the MDL code every example
consists of two reals which are encoded as 64 bits.
Thus, the erroroneous examples, those exceeding the small
fixed error cut-off level that we set, are encoded
in 64 bits each. We ignore the amount with which the misclassified output
real value differs from the target real value, it may be large or small.
The total error is encoded as an explicit list 
of the misclassified
examples.

Because MDL selects the model
that minimizes sum of model length and total
error length, 
it is important how large a training set we choose. 
The coding length of the models is the same for
every fixed $k$ and training set size, but the total error length depends
on this. For a small training set the number of erroneous (misclassified)
examples may be very small compared to the model code length, and
the difference between simple models with small $k$ and complex models
is large. With large training sets the opposite happens. This is exactly
right: with a small number of examples the simpler models are
encouraged. How complex a model can be must be justified by
the size of the training set. Intuitively, with increasing training set size,
eventually the smallest model
that has low error on this set can be expected to stabilize and
to have low prediction error.

\begin{figure}[htb]
\begin{center}
\leavevmode
\def\epsfsize#1#2{0.33#1}
\vspace*{5mm}
\epsfbox{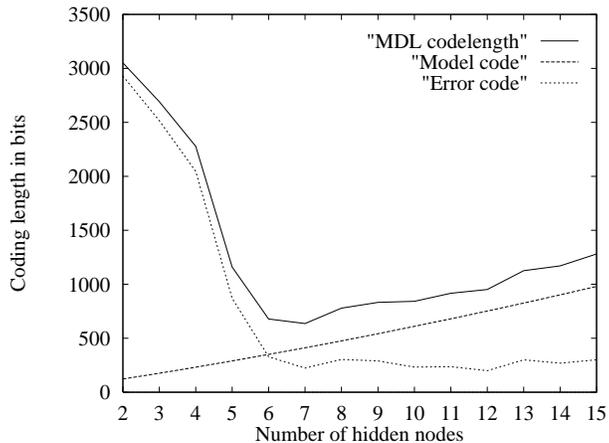}
\vspace*{-2mm}
\caption{\label{test:draw}Prediction by MDL.}
\end{center}
\vspace*{-8mm}
\end{figure}

In the following experiment
we used a random training set of 100 examples, and
for every $k$ ($2 \leq k \leq 15$) a network with $k$ nodes in the hidden layer 
was trained with $10^6$ training cycles.
Figure~\ref{test:draw} shows the results in terms of MDL:
the model code length, the error code length, and the
total description length, as a function of the number
of nodes in the hidden layer. The optmimum of the total code
length is reached for seven hidden nodes, that is, MDL predicts that
seven hidden nodes (the granularity of the hidden layer so to speak)
give the best model.
This is only one node away from the optimal network size determined 
experimentally below.

\subsection{Validation of the Model}

\begin{figure}[hbt]
\vspace*{-10mm}
\begin{center}
\leavevmode
\flushleft
\hspace*{180pt}
\def\epsfsize#1#2{0.23#1}
\epsfbox{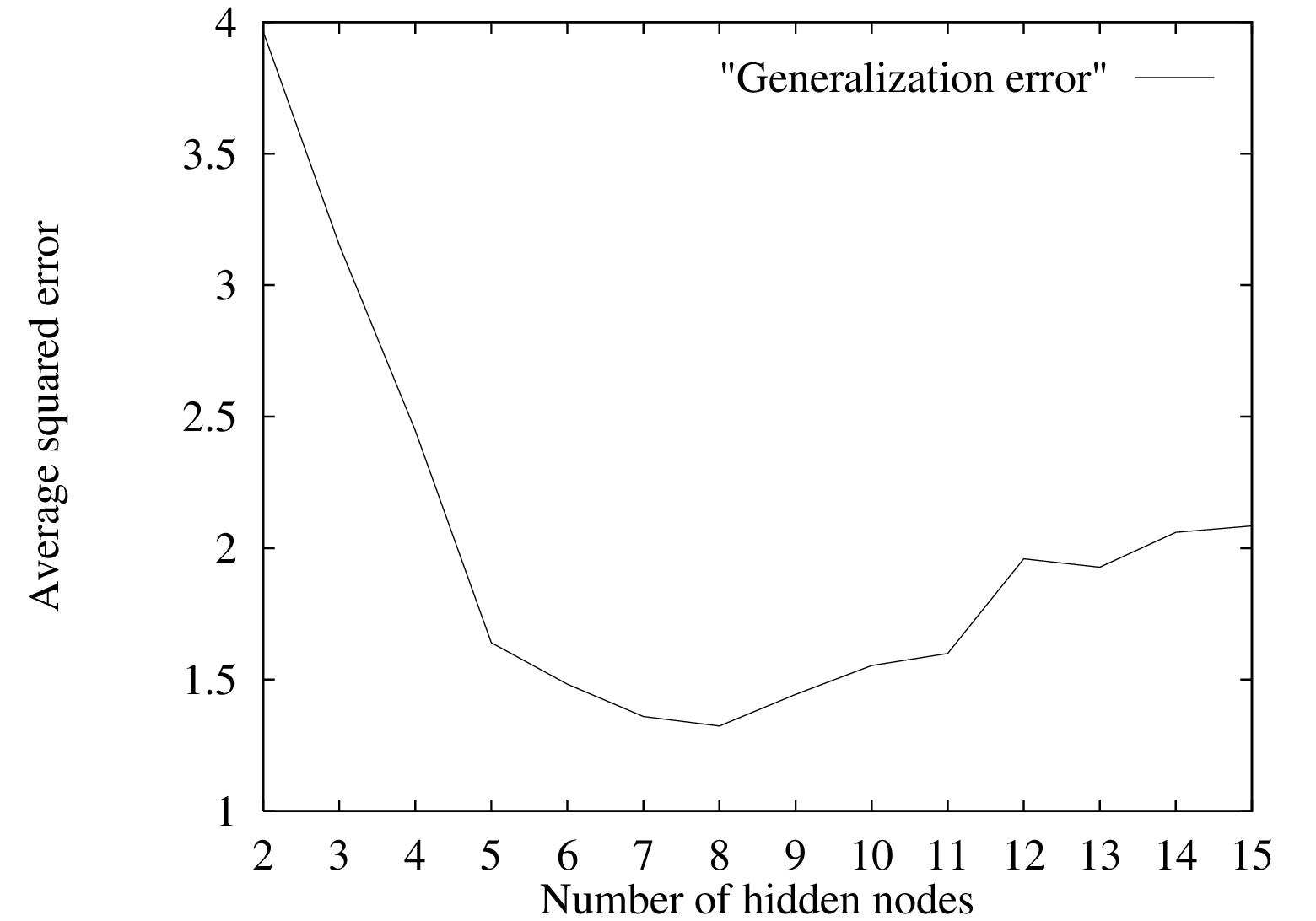}
\vspace*{-125pt}
\flushleft
\hspace*{0pt}
\epsfbox{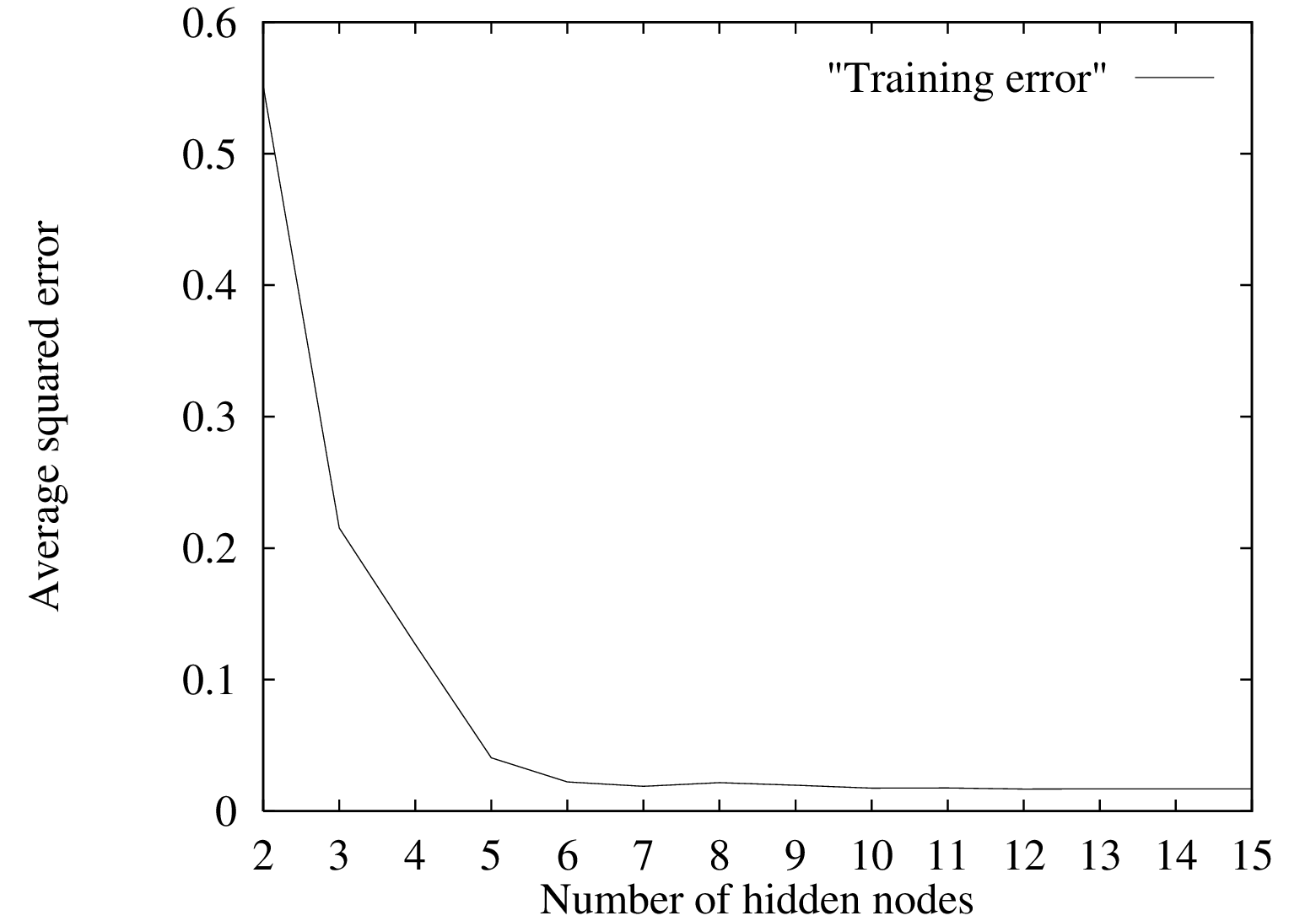}
\vspace*{-2mm}
\caption{\label{err:draw}Error on the training set and the test set}
\end{center}
\vspace{-5mm}
\end{figure}

To determine the best number of hidden nodes we used 
thirty different random training sets of 100 examples each.
For every $k$ ($2 \leq k \leq 15$) the network
was trained using $10^6$ training cycles.
Other more sophisticated stop criteria could have been 
used, but some checking showed that in general the performance
of the network after $10^6$ training cycles was close to optimal.
The error per example is
the real distance between the output vector and the target vector.
In figure~\ref{err:draw} the average squared error on the training set 
and the average squared prediction error on the unseen test set
are displayed as a function of the number of nodes in the
hidden layer. 
The optimal network in the sense of having best
extrapolation and generalization properties in
modeling the unseen examples in the test set most correctly,
is a network with 8 hidden nodes.

As expected, the error on the training set keeps on decreasing
with an increasing number of nodes in the hidden layer, that is,
when the model becomes increasingly complex and is capable of
modeling more and more detail.
When we look at the prediction error of new examples that were not in the
training data (figure~\ref{err:draw}), we see that the average squared
prediction error first
decreases when the model complexity increases, but that there is an
optimum of  minimum error
after which the error starts to increase again.  
Experimentally, the best number of hidden nodes for
this problem with a training set size of 100 examples is 8, that is,
one more than predicted by the simplified application of MDL above.

\section{Discussion}
\par\noindent
We applied the theoretical Minimum Description Length principle 
to two different  experimental tasks aimed at learning the best
model discretization. The first application was learning 
to recognize isolated handwritten characters
on-line using elastic matching
and some statistical technique. A `model' is a collection of
prototypes built from raw training character drawings
by on-line taking points
on the curves of the executed character drawing at a constant feature
extraction interval, and by combining closely related character drawings.
Some novel features here are the use of multiple prototypes
per character, and the use of the MDL principle to choose
the optimal feature extraction interval.

The model is optimized in the spirit of MDL by minimizing the total
code length, which is the sum of the model and error-to-model code
lengths, against different feature extraction intervals. The resulting model is
optimal according to the theory. 
It is then validated by
testing using a different set of character drawings from the same source.
We believe that the result of this small test gives evidence 
that MDL may be a good tool in 
the area of handwritten character recognition.

The second application was
modeling a robot arm by a three layer
feedforward neural network, where the precision parameter to be learned
is the number of nodes in the hidden layer. The MDL predicted number
of nodes was validated by extensive testing of the model with
respect to extrapolation and generalization capabilities using
unseen examples from a test set.

The optimal granularity of the models was predicted for
sensible values--only marginally different from the 
experimentally determined optimal ones. This shows that this rigorous and
not ad hoc form of
``Occam's Razor'' is quite succesfull.
Comparison of the performance of the---admittedly limited---
experiments on the robot arm problem
with that of other principles, such as
NIC and AIC, indicated that MDL's performance was better or competitive
\cite{BKV94}.

A similar theory and practice validation in case of the 
Bayesian framework for model comparison
was given by Mackay \cite{mckay92b}. This paper inspired
us to use the robot arm problem in the MDL setting. We note
that the Bayesian framework is genuinely different as is rigorously
demonstrated in our companion paper \cite{ViLi99}. 
It is well known that 
prefix code length is equivalent
to negative log probability through the Shannon-Fano code \cite{CT91,LiVibook},
and therefore with every such code there corresponds an equivalent 
probability. Thus, the MDL coding approach can in principle be
translated back into a Bayesian approach where the model code 
gives the prior. The analysis we have given in \cite{ViLi99} shows that the
data-to-model error may not correspond the the conditional data-to-model
probability if the data are ``atypical'' for the contemplated prior.
Moreover, to the authors coding of large data is more natural
than reasoning about possibly nonexisting probabilities.

\subsection{Directions for Future Work}
In general the MDL method appears to be well
suited for supervised learning of best model discretization parameters
for classification problems in which
error coding is straightforward.
Applying the MDL method is
simple, and it is computationally not expensive.

The central point is that using MDL 
the optimal granularity of the model parameters
can be computed automatically rather than tuned manually.
This approach constitutes a rational and feasibly computable approach for
feature selection as opposed to customary rather ad hoc approaches. 
The purpose of presenting the theory outline and
 the example applications is to stimulate
re-use in different areas
of pattern recognition, classification, and image understanding
(region segmentation, color clustering segmentation, and so on).

\section*{Acknowledgement}
\par\noindent
We are grateful to Les Valiant for many discussions
on machine learning and the suggestion for this research,
to Guido te Brake and Joost Kok for executing the robot arm
experiment, and to the referees for their insightful comments.

\appendix
\section{Appendix: Kolmogorov Complexity}
\label{app.A}
\label{sect.kc}
The Kolmogorov complexity \cite{Ko65} of a finite object $x$
is the length of the
shortest effective binary description of $x$.
We give a brief outline of definitions and properties.
For more details see \cite{LiVibook}.
Let $x,y,z \in {\cal N}$, where
${\cal N}$ denotes the natural
numbers and we identify
${\cal N}$ and $\{0,1\}^*$ according to the
correspondence
\[(0, \epsilon ), (1,0), (2,1), (3,00), (4,01), \ldots \]
Here $\epsilon$ denotes the {\em empty word} `' with no letters.
The {\em length} $l(x)$ of $x$ is the number of bits
in the binary string $x$. For example,
$l(010)=3$ and $l(\epsilon)=0$.

The emphasis is on binary sequences only for convenience;
observations in any alphabet can be so encoded in a way
that is `theory neutral'.

A binary string $x$
is a {\em proper prefix} of a binary string $y$
if we can write $x=yz$ for $z \neq \epsilon$.
 A set $\{x,y, \ldots \} \subseteq \{0,1\}^*$
is {\em prefix-free} if for any pair of distinct
elements in the set neither is a proper prefix of the other.
A prefix-free set is also called a {\em prefix code}.
Each binary string $x=x_1 x_2 \ldots x_n$ has a
special type of prefix code, called a
{\em self-delimiting code},
\[ \bar x =x_1x_1x_2x_2 \ldots x_n \neg x_n ,\]
where
$\neg x_n=0$ if $x_n=1$ and $\neg x_n=1$ otherwise. This code
is self-delimiting because we can determine where the
code word $\bar x$ ends by reading it from left to right without
backing up. Using this code we define
the standard self-delimiting code for $x$ to be
$x'=\overline{l(x)}x$. It is easy to check that
$l(\bar x ) = 2 n$ and $l(x')=n+2 \log n$.

We develop the theory using 
Turing machines, but we can as well use the 
set of LISP programs or the set of FORTRAN programs.

Let $T_1 ,T_2 , \ldots$ be a standard enumeration
of all Turing machines, and let $\phi_1 , \phi_2 , \ldots$
be the enumeration of corresponding functions
which are computed by the respective Turing machines.
That is, $T_i$ computes $\phi_i$.
These functions are the {\em partial recursive} functions
or {\em computable} functions. The Kolmogorov complexity
$C(x)$ of $x$ is the length of the shortest binary program
from which $x$ is computed. Formally, we define this as follows.

\begin{definition}\label{def.KolmC}
The {\em Kolmogorov complexity} of $x$ given $y$ (for
free on a special input tape) is
\[C(x|y) = \min_{p,i}\{l(i'p): \phi_i (p,y )=x , p \in \{0,1\}^*, i
\in {\cal N} \}. \]
Define $C(x)=C(x|\epsilon)$.
\end{definition}

Though defined in terms of a
particular machine model, the Kolmogorov complexity
is machine-independent up to an additive
constant
 and acquires an asymptotically universal and absolute character
through Church's thesis, from the ability of universal machines to
simulate one another and execute any effective process.
  The Kolmogorov complexity of an object can be viewed as an absolute
and objective quantification of the amount of information in it.
   This leads to a theory of {\em absolute} information {\em contents}
of {\em individual} objects in contrast to classic information theory
which deals with {\em average} information {\em to communicate}
objects produced by a {\em random source} \cite{LiVibook}.

For technical reasons we also need a variant of complexity,
so-called prefix kolmogorov complexity, which is associated with Turing machines
for which the set of programs resulting in a halting computation
is prefix free. We can realize this by equiping the Turing
machine with a one-way input tape, a separate work tape,
and a one-way output tape. Such Turing
machines are called prefix machines
since the halting programs for anyone of them form a prefix free set.
Taking the universal prefix machine $U$ we can define
the prefix complexity analogously with the plain Kolmogorov complexity.
If $x^*$ is the first shortest program for $x$ then the set
$\{x^* : U(x^*)=x, x \in \{0,1\}^*\}$ is a {\em prefix code}.
That is, each $x^*$ is a code word for some $x$, and if $x^*$
and $y^*$ are code words for $x$ and $y$ with $x \neq y$ then $x^*$ is not
a prefix of $x$.

Let $\langle \cdot \rangle$ be a standard invertible
effective one-one encoding from ${\cal N} \times {\cal N}$
to prefix-free recursive subset of ${\cal N}$.
For example, we can set $\langle x,y \rangle = x'y'$.
We insist on prefix-freeness and
recursiveness because we want a universal Turing
machine to be able to read an image under $\langle \cdot \rangle$
from left to right and
determine where it ends.

\begin{definition}\label{def.KolmK}
The {\em prefix Kolmogorov complexity} of $x$ given $y$ (for
free) is
\[K(x|y) = \min_{p,i}\{l(\langle p,i\rangle): \phi_i (\langle p,y \rangle )=x ,
p \in \{0,1\}^*, i
\in {\cal N} \}. \]
Define $K(x)=K(x|\epsilon)$.
\end{definition}

The nice thing about $K(x)$ is that we can interpret $2^{-K(x)}$
as a probability distribution since $K(x)$ is the length of
a shortest prefix-free program for $x$. By the fundamental
Kraft's inequality, see for example \cite{CT91,LiVibook}, we know that
if $l_1 , l_2 , \ldots$ are the code-word lengths of a  prefix code,
then $\sum_x 2^{-l_x} \leq 1$. This leads to the notion
of algorithmic universal distribution---a rigorous form of Occam's razor--below.

\section{Appendix: Universal Distribution}
\label{app.B}
A Turing machine $T$ computes a function on the natural numbers.
However, we can also consider the computation
of real valued functions. For this purpose we consider
both the argument of $\phi$ and the value of $\phi$
as a pair of natural numbers according to the standard
pairing function $\langle \cdot \rangle$. We define
a function from ${\cal N}$ to the reals ${\cal R}$
by a Turing machine $T$ computing
a function $\phi$ as follows. Interprete
the computation $\phi(\langle x,t \rangle ) = \langle p,q \rangle$
to mean that the quotient $p/q$ is
the rational valued $t$th approxmation of $f(x)$.

\begin{definition}\label{def.enum.funct}
A function $f: {\cal N} \rightarrow {\cal R}$ is
{\em enumerable} if there is a Turing machine $T$ computing a
total function $\phi$
such that $\phi (x,t+1) \geq \phi (x,t)$ and
$\lim_{t \rightarrow \infty} \phi (x,t)=f(x)$. This means
that $f$ can be computably approximated from below.
If $f$ can also be computably approximated from above
then we call $f$ {\em recursive}.
\end{definition}

A function $P: {\cal N} \rightarrow [0,1]$ is
a {\em probability distribution} if
$\sum_{x \in {\cal N}} P(x) \leq 1$. (The inequality
is a technical convenience. We can consider
the surplus probability to be concentrated on the
undefined element $u \not\in {\cal N}$).

Consider the family ${\cal EP}$ of
{\it enumerable} probability distributions on the
sample space ${\cal N}$ (equivalently, $\{0,1\}^*$).
It is known, \cite{LiVibook}, that ${\cal EP}$
contains an element $\hbox{\bf m}$ that
multiplicatively dominates all elements of ${\cal EP}$. That is,
for each $P \in {\cal EP}$ there is a constant $c$ such
that $c \: \hbox{\bf m} (x) > P(x)$ for all $x \in {\cal N}$.
We call ${\bf m}$ an 
{\em algorithmic universal distribution}
or shortly {\em universal distribution}.

The family ${\cal EP}$ contains all distributions
with computable parameters
which have a name, or in which we could conceivably
be interested, or which have ever been considered.
The dominating property means that $\hbox{\bf m}$
assigns at least as much probability to each object
as any other distribution in the family ${\cal EP}$
does. In this sense it is a universal {\em a priori}
by accounting for maximal ignorance. It turns out that
if the true {\em a priori} distribution in Bayes's rule
is recursive, then using the single distribution
$\hbox{\bf m}$, or its continuous analogue the measure $\hbox{\bf M}$
on the sample space $\{0,1\}^{\infty}$ (for prediction as in
\cite{So78})
is provably as good
as using the true {\em a priori} distribution.

We also know, \cite{LiVibook}, that we can choose
\begin{equation}\label{eq.m}
 - \log \hbox{\bf m} (x) = K(x) 
\end{equation}
That means that $\hbox{\bf m}$ assigns high probability to simple
objects
and low probability to complex or random objects.
For example, for $x=00 \ldots 0$ ($n$ 0's) we have
$K(x) \eqa K(n) \lea \log n + 2 \log \log n $ since the program
\[ \mbox{\tt print } n \mbox{\tt \_times a ``0''} \]
prints $x$. (The additional $2 \log \log n$ term
is the penalty term for a self-delimiting encoding.)
Then, $1/ (n \log^2 n ) = O( \hbox{\bf m}(x))$.
But if we flip a coin to obtain a string $y$ of $n$ bits,
then with overwhelming probability $K(y) \gea n $
(because $y$ does not contain effective regularities
which allow compression),
and hence $\hbox{\bf m}(y) = O( 1/2^n)$.

The algorithmic
universal distribution has many astonishing properties \cite{LiVibook}.
One of these, of interest to the AI community, is that it gives
a rigorous meaning to Occam's Razor by assigning high probability
to the ``simple,'' ``regular,'' objects and low probability
to the ``complex,'' ``irregular'', ones. For a popular account see
\cite{KLV97}.
 A celebrated result states that
an object $x$ is individually random
with respect to a conditional probability distribution $P(\cdot |y)$
iff $\log ({\bf m}(x|y)/P(x|y)) \eqa 0$. Here the implied
constant in the $\eqa$ notation is in fact related
to $K(P(\cdot |y))$---the length of the shortest program
that computes the probability $P(x|y)$ on input $x$. In particular
this means that for $x$ is ``typical'' or ``in general position''
with respect to conditional distribution $P( \cdot |y)$ iff
the real probability $P(x|y)$ is close to the
algorithmic universal probability ${\bf m}(x|y)= 2^{- K(x|y)}$.

\bibliographystyle{plain}

\begin{thebibliography}{10}

\bibitem{BRY98}
A.R. Barron, J. Rissanen, and B. Yu,
The minimum description length principle in coding and modelling,
{\em IEEE Trans.\ Inform.\ Theory},
IT-44:6(1998), 2743--2760.

\bibitem{Bi95}
C. Bishop,
{\em Neural Networks for Pattern Recognition}, Oxford University Press,
1995.


\bibitem{BKV94}
G. te Brake, J.N. Kok, and P.M.B. Vit\'anyi,
Model selection for neural networks: Comparing MDL and NIC,
In: {\em Proc.  European
Symposium on Artificial Neural Networks}, Brussels, April 20-22, 1994.

\bibitem{CT91}
T.M. Cover and J.A. Thomas, {\em Elements of Information Theory},
Wiley, New York, 1991.


\bibitem{GL89}
Q.~Gao and M.~Li.
\newblock An application of minimum description length principle to online
  recognition of handprinted alphanumerals.
\newblock In {\em 11th International Joint Conference on Artificial
  Intelligence}, pages 843--848. Morgan Kaufmann Publishers, 1989.




\bibitem{Je61}
H.~Jeffreys.
\newblock {\em Theory of Probability}.
\newblock Oxford at the Clarendon Press, Oxford, 1961.
\newblock Third Edition.

\bibitem{KV}
M.J. Kearns, U. Vazirani,
{\em An Introduction to Computational Learning Theory},
MIT Press, 1994.

\bibitem{KLV97}
W.W. Kirchherr, M. Li and P.M.B. Vit\'anyi,
The miraculous universal distribution,
{\em Mathematical Intelligencer}, 19:4(1997), 7--15.


\bibitem{Ko65}
A.N. Kolmogorov.
\newblock Three approaches to the quantitative definition of information.
\newblock {\em Problems Inform. Transmission}, 1(1):1--7, 1965.

\bibitem{Ku87}
J.~Kurtzberg.
\newblock Feature analysis for symbol recognition by elastic matching.
\newblock {\em IBM J. Res. Develop.}, 31(1):91--95, 1987.


\bibitem{LiVi89a}
M.~Li and P.M.B. Vit\'anyi.
\newblock Inductive reasoning and {Kolmogorov} complexity.
\newblock {\em J. Comput. Syst. Sci.}, 44:343--384, 1992.

\bibitem{LiVibook}
M.~Li and P.M.B. Vit\'anyi.
{\em An Introduction to {K}olmogorov Complexity and Its
  Applications}, Springer-Verlag, New York, 2nd Edition, 1997.

\bibitem{mckay92b}
D.J.C. MacKay.
\newblock A practical bayesian framework for backpropagation networks.
\newblock {\em Neural Computation}, 4(3):448--472, 1992.

\bibitem{ML66}
P. Martin-L\"of, The definition of random sequences,
{\em Inform. Contr.}, 9(1966), 602-619.


\bibitem{Mi19}
R. von Mises, Grundlagen der {W}ahrscheinlichkeitsrechnung,
{\em Mathemat. Zeitsch.}, 5(1919), 52-99.



\bibitem{QuRi89}
J.~Quinlan and R.~Rivest.
\newblock Inferring decision trees using the minimum description length
  principle.
\newblock {\em Inform. Computation}, 80:227--248, 1989.

\bibitem{Ri78}
J.~Rissanen.
\newblock Modeling by the shortest data description.
\newblock {\em Automatica-J.IFAC}, 14:465--471, 1978.

\bibitem{Ri84}
J.~Rissanen.
\newblock Universal coding, information, prediction and estimation.
\newblock {\em IEEE Transactions on Information Theory}, IT-30:629--636, 1984.

\bibitem{Ri86}
J.~Rissanen.
\newblock Minimum description length principle.
\newblock In S.~Kotz and N.L. Johnson, editors, {\em Encyclopaedia of
  Statistical Sciences, Vol. V}, pages 523--527. Wiley, New York, 1986.

\bibitem{Ri87a}
J.~Rissanen.
\newblock Stochastic complexity.
\newblock {\em J. Royal Stat. Soc., series B}, 49:223--239, 1987.
\newblock Discussion: pages 252-265.

\bibitem{So64}
R.J. Solomonoff, 
A formal theory of inductive inference, Part 1 and Part 2,
{\em Inform. Contr.}, 7(1964), 1-22, 224-254.

\bibitem{So78}
R.J. Solomonoff.
\newblock Complexity-based induction systems: comparisons and convergence
  theorems.
\newblock {\em IEEE Trans. Inform. Theory}, IT-24:422--432, 1978.

\bibitem{Su80}
C.Y. Suen.
\newblock Automatic recognition of handprinted characters---the state of art.
\newblock {\em Proc. IEEE}, 68(4):469--487, 1980.

\bibitem{Ta82}
C.C. Tapper.
\newblock Cursive script recognition by elastic matching.
\newblock {\em IBM J. Res. Develop.}, 26(6):765--771, 1982.

\bibitem{Va84a}
L.G. Valiant.
\newblock Deductive learning.
\newblock {\em Phil. Trans. Royal Soc. Lond.}, A 312:441--446, 1984.

\bibitem{Va84}
L.G. Valiant.
\newblock A theory of the learnable.
\newblock {\em Comm. Assoc. Comput. Mach}, 27:1134--1142, 1984.

\bibitem{ViLi99}
P.M.B. Vit\'anyi and M. Li,
Minimum Description Length Induction,  Bayesianism,
and Kolmogorov Complexity,
{\em IEEE Trans. Inform. Theory}, IT-46:2(2000), 446--464.


\end{thebibliography}

\end{document}